\documentclass[preprint,showpacs,amsmath,amssymb]{revtex4}

\usepackage{graphicx}% Include figure files
\usepackage{dcolumn}% Align table columns on decimal point
\usepackage{bm}% bold math
\usepackage{anysize}
\usepackage{graphicx}
\usepackage{caption}
\usepackage{graphicx}% Include figure files
\usepackage{dcolumn}% Align table columns on decimal point
\usepackage{bm}% bold math
\usepackage{amsmath}
\usepackage{amssymb}
\usepackage{mathrsfs}
\usepackage[latin1]{inputenc}
\usepackage{amsmath}

\newcommand{\p}{\partial}

\newcommand{\e}{\epsilon}
\newcommand{\g}{\mathbf{g}}

\newcommand{\sd}{Schr\"{o}dinger }

\newcommand{\tr}{{\rm Tr}}

\newcommand{\U}{\mathcal{U}}
\newcommand{\G}{\mathcal{G}}
\newcommand{\SU}{\mathcal{SU}}

\newcommand{\D}{\mathcal{D}}
\newcommand{\M}{\mathcal{M}}

\newtheorem{theorem}{Theorem}

\begin{document}

%\preprint{APS/123-QED}

%\title{Quantum Control Landscapes with $\SU(2)$ Dynamic Symmetry}
\title{The role of controllability in optimizing quantum dynamics}

\author{Re-Bing Wu}
\affiliation{Department of Automation, Tsinghua University, Beijing,
100084, China\\
Center for Quantum Information Science and Technology, TNList,
Beijing, 100084, China.} \email{rbwu@tsinghua.edu.cn}
\author{Michael A. Hsieh}
\affiliation{Department of Chemistry and Center for Quantum Information Science and Technology, University of
Southern California, Los Angeles, CA 90025, USA}

\author{Herschel Rabitz}
\altaffiliation{Department of Chemistry, Princeton University,
Princeton, New Jersey 08544, USA}

\date{\today}% It is always \today, today,

\begin{abstract}
This paper reveals an important role that controllability plays in reducing the complexity of optimizing quantum control dynamics. Based on the quantum control landscape, which refers to the optimization
objective as a functional of the control fields, we analyze the critical topology of the gate fidelity landscape for uncontrollable systems due to $\mathcal{SU}(2)$ and $\mathcal{SU}(3)$ dynamical symmetries. The loss of controllability leads to multiple local suboptima on the control landscape, which form traps on the way towards seeking the global optimal value for the objective. The control landscape generally becomes rugged when controllability is constrained by reduced dynamical symmetry, and even more rugged if in addition the target gate is not realizable. These results imply that the optimization of quantum dynamics can be seriously impeded by local search algorithms under these conditions.
\end{abstract}

%\pacs{03.67.-a,02.20.Qs,02.20.-a  }% PACS, the Physics and Astronomy
%\keywords{Quantum control, recurrent, scattering}%Use showkeys class option if keyword
\maketitle

\section{Introduction}
A quantum system is referred to as controllable if it can be transferred from an arbitrary state to another through engineering the dynamics. As a fundamental aspect of quantum control theory, controllability has been widely studied for various types of systems including infinite-dimensional\cite{Huang1983,WuTarn2006} and finite-dimensional\cite{RamRab1995} closed systems and Markovian open systems\cite{Altafini2003}. However, the practical necessity for controllability is still unclear as it may not be required in many realistic applications and may not always be available. For example, the goal of realizing only a specific unitary transformation related to achieving a quantum computation algorithm, or reaching particular excited
states of a molecule generally do not require full controllability. No apparent advantage can be seen by merely gaining additional controllability. The study of this issue is also motivated by the growing number of successful adaptive control experiments utilizing shaped laser pulses as the controls \cite{Brif2010NJP}. In most of these systems, there is no assessment of controllability and a simple yes-or-no controllability answer, likely has little practical relevance.

Here we study this matter in the framework of the quantum control
landscape \cite{RabMik2004}, which was developed for understanding the
complexity of optimizing quantum dynamics
\cite{Rabitz2000,Brif2010NJP}. A quantum control landscape
is formally defined as a target functional, $J[\{\epsilon_k(\cdot)\}]$, to be maximized (minimized) over the set of implementable control fields $\{\e_1(t),\cdots,\e_m(t)\}$ that steer the following $N$-level quantum control system:
\begin{equation}\label{quantum control system}
    \frac{\p U(t)}{\p t}=(i\hbar)^{-1}\left\{H_0+\sum_{k=1}^m\e_k(t)H_k\right\}U(t),~~U(0)=\mathbb{I},
\end{equation}
where $U(t)\in\U(N)$ is the system propagator at time $t$, $H_0$ is the internal
Hamiltonian and $\{H_1,\cdots,H_m\}$ are the control Hamiltonians.

A major goal of quantum control landscape theory is performance of a topological analysis of the critical points (i.e., the critical topology) from which insights can be gained about the complexity of optimizing quantum system dynamics. Consider the class of control landscapes in the form of $J=J(U(t_f))$, where the system
propagator $U(t_f)$ at $t=t_f$ is an implicit functional of the
control field through the \sd equation (\ref{quantum control system}). The control landscape can be displayed either on the space of the control fields (i.e., in the dynamical picture) or on the set of realizable propagators $U(t_f)$ (i.e., in the kinematic picture). In prior studies \cite{WuPech2008,WuDomi2008}, we have shown that the main topological characteristics of the critical points in the dynamical
picture are preserved when mapped onto the critical points in the kinematic picture except for what appear to uncommon singularity issues \footnote{there is a special class of so called singular controls that behaves differently, but their influence generally does not appear to be important on the quantum control landscape topology \cite{WuDomi2008}}. Thus, almost all maxima, minima and saddle points in the dynamical picture should correspond to the same type of critical points in the easier to analyze kinematic picture without distortion of the topological features.

The kinematic control landscape may be constructed from the (low-dimensional) set of realizable propagators. If the system is controllable, then the set of realizable propagators form the unitary group. This circumstance has been studied for practical optimization objectives (e.g., the state transfer probability \cite{RabMik2004}, ensemble average \cite{WuRabi2008} and gate fidelity \cite{Rabitz2005}), which collectively show that none of these landscapes possess local suboptima that may impede the search for the ultimate global optima. This behavior provides a basis to explain the large and growing body of experimental quantum control successes, although in the laboratory many ancillary issues and constraints may be present.

Realistic quantum systems may not be always fully manipulable. A typical circumstance is when the system is under some dynamic symmetry that restricts the system's evolution. Even if the system is not subject to a dynamic symmetry, the restrictions on available control resources (e.g., the amplitude, bandwidth and center frequency of the applied field) can still cause a practical loss of full controllability, in which case movement over the landscape when searching for an optimal control can become limited on the otherwise trap free landscape. Whether these uncontrollable systems still have a physically attractive landscape topology becomes practically important for assessing the complexity of searching for optimal controls, which is explored in this paper.

The loss of controllability poses restrictions on the set of reachable states, which impacts the kinematic analysis. The set of reachable states for control-restricted systems is generally very complex to identify, while that for systems with dynamic symmetry may be simply revealed (e.g., as a Lie subgroup of $\U(N)$). The landscape topology for systems with dynamic symmetry has been identified for gate fidelity under certain special symmetries \cite{WuRaj2010,HsiehWu2010}, where no false traps were observed. However, as will be seen in this paper, false traps (i.e., yet appearing as real in practice) can occur in the landscape of symmetry induced uncontrollable quantum systems. The paper will be organized as follows. Section II discusses the general conditions for the existence of critical points on such landscapes. Section III analyzes the case of the fidelity control landscape of transformations with $\SU(n)$ dynamic symmetry, where for $n=2$ or $n=3$ we derive the analytic structure for simple cases and graphically compare the landscapes for different symmetry types. Finally, conclusions are drawn in Section IV.

\section{general conditions for critical points on a landscape with dynamic symmetry}
Let $\g$ be the Lie algebra spanned by
$(i\hbar)^{-1}H_0,(i\hbar)^{-1}H_1,\ldots,(i\hbar)^{-1}H_m$, then the quantum propagator must evolve within the Lie group $\G=\exp \g$ generated by $\g$, which forms the dynamic symmetry of the system. If $\g$ forms a proper Lie subalgebra of
$u(N)$, the system must be uncontrollable. However, owing to the compactness of the unitary group, the set of realizable propagators can fill up the group $\G$ if the control is not limited and the evolution time is sufficiently long. Hence, in a kinematic analysis, one can suitably study the landscape of symmetry induced uncontrollable quantum systems on the corresponding symmetry group.

The condition for $\tilde U\in \G$ to be a landscape critical point can be derived through parameterizing its neighborhood in $\G$ as $\tilde Ue^{A}$, where $A\in\g$ represents the local coordinates. The variation of the landscape function with respect to $A$ can then be written via the chain rule
\begin{equation}\label{chain rule}
\delta J_\G=\left\langle \nabla J(\tilde U),\tilde U\delta
A\right\rangle=\left\langle \tilde U^\dag\nabla J(\tilde U),\delta
A\right\rangle\equiv 0,~~~\forall~\delta A\in \g,
\end{equation}where $\nabla J(U)$ is the gradient function over $\U(N)$ and
the matrix inner product $\langle X,Y\rangle=Re\tr(X^\dag Y)$. Let $\mathbb{D}(\tilde U)=\tilde U^\dag\nabla J(\tilde U)\in u(N)$,
then the
condition for $\tilde U\in \G$ to be critical is that
\begin{equation}\label{projection}
\langle\mathbb{D}(\tilde U),\g\rangle=0,
\end{equation}
which requires that the vector $\mathbb{D}(\tilde U)\in {\bf \rm u}(N)$ be orthogonal to the subspace $\g$. For example, $\mathbb{D}(U)=[U\rho U^\dag ,\theta ]$ corresponds to the ensemble control landscape $J=\tr(U\rho U^\dag \theta)$ and $\mathbb{D}(U)=W^\dag U-U^\dag W$ corresponds to the gate fidelity control landscape $J=Re\tr(W^\dag U)$.
The set of critical points with $\G$ dynamic symmetry can then be expressed as
$$\M_\G=\{U\in\G:~\langle\mathbb{D}(U),\g\rangle=0.\}$$

The condition (\ref{projection}) provides a clear geometric picture for the critical points, however, it is difficult to use for directly determining the critical topology with an arbitrary dynamic symmetry, except in the extreme case that the system is controllable (i.e., $\g=\mathbf{u}(N)$), where the condition
can be simplified to $\mathbb{D}(\tilde U)=0$ on $\U(N)$. It is also possible, as will be shown below, to simplify the analysis by making use of the symmetric form of the landscape function, from which we can either solve for the critical topology or easily visualize it for low dimensional symmetries.

\section{Quantum Gate fidelity Control landscape with $\SU(n)$ dynamic symmetry}
The quantum gate fidelity landscape is defined as
\begin{equation}\label{W}
J(U)=N^{-1}|\tr(W^\dag U)|
\end{equation}
where $N$ is the dimension of the system and $W\in\U(N)$ is the
target unitary transformation \footnote{Note that this function is
similar, but distinct from the one studied in \cite{Rabitz2005}. We
use the present one to avoid the influence of the trivial global phase. One
can prove that, although more complicated, the landscape for $J$ in Eq. \ref{W} is still
trap-free for controllable systems \cite{Ho2009}.}. If the target transformation $W$ is realizable, i.e., $W\in \G$, then we can use
the transformation $\tilde U= W^\dag U$ to change the landscape function
into a canonical form $J_\G(\tilde U)=N^{-1}|\tr(\tilde U)|$.

From group representation theory, a dynamic symmetry group $\G\subset \U(N)$ can be taken as a $N$-dimensional unitary representation of the abstract Lie group $\G$, which thereby can be decomposed into smaller unitary irreducible representations (UIR), i.e., one can always
find a constant unitary similarity transformation in $\G$ that turns each
$U\in\G$ into a block diagonal form:
$$U=\left(
    \begin{array}{ccc}
      U^{(1)} &  &  \\
       & \ddots &  \\
       &  & U^{(r)} \\
    \end{array}
  \right),
$$where $U^{(j)}\in\U(N_j),~~N_1+\cdots+N_r=N$, and each sub-block $U^{(j)}$ cannot be decomposed into smaller ones.
For the gate fidelity control landscape, the landscape function with $\G$ carrying a reducible representation can be written as the sum of the landscape function values evaluated on its irreducible components. Hence, it is convenient to start with the case that $\G$ is a unitary irreducible representation of some abstract Lie group in $\U(N)$.

A classification scheme of uncontrollable systems was proposed \cite{Polack2009} based on the Lie algebra $\g$ and their unitary irreducible representations (UIR). Here we only consider the class of $\SU(n)$ dynamic symmetry (i.e., the group of $n\times n$ unit trace unitary matrices). For example, the $\SU(2)$ dynamic symmetry may arise when the free and control
Hamiltonians of spin-$j$ particles (or a rigid rotor whose angular
momentum is $j^2$) in a magnetic field are a linear combination of the
spin operators $J_x$, $J_y$ and $J_z$ that possess an
$N=2j+1$ dimensional UIR (denoted by $\mathcal{D}_j$). Similarly, two non-interacting spin
particles with spin-$j_1$ and spin-$j_2$ in a magnetic field (e.g.,
two electrons in a multi-electron molecular system) possess a tensor
product of $\SU(2)$ UIRs $\mathcal{D}=\mathcal{D}_{j_1}\otimes
\mathcal{D}_{j_2}$, which can be decomposed into the direct sum of smaller UIRs of $\SU(2)$.

Suppose that $\G$ carries an $N$-dimensional UIR of $\SU(n)$, then there must be a homeomorphic mapping $\mathcal{R}$ that transforms any unitary matrix $U\in \G\subset \U(N)$ to an $n\times n$ unitary matrix $U'\in \SU(n)$ such that $U=\mathcal{R}(U')$. Let the complex numbers $\e_1,\cdots,\e_n$ be the eigenvalues of $U'$ corresponding to $U$, which satisfy $|\e_1|=\cdots=|\e_n|=1=\e_1\cdots \e_n$, then according to the well-known Weyl formula \cite{PLYMEN1976}, the UIRs of $\SU(n)$ are labeled by a group of integers $r_1>\cdots>r_{n-1}$ \cite{PLYMEN1976}, and the trace of $U$ can be expressed as the following character function
\begin{eqnarray*}\label{xi} \tr(U)&=&\chi_{r_1,\cdots,r_{n-1}}(\e_1,\cdots,\e_n)\\
&=&|\e^{r_1},\cdots,\e^{r_{n-1}},1|/|\e^{n-1},\cdots,\e,1|,
\end{eqnarray*}
where $$|\e^{r_1},\cdots,\e^{r_{n-1}},1|={\rm det}\left(
                                \begin{array}{cccc}
                                  \e_1^{r_1} & \cdots & \e_1^{r_{n-1}} & 1 \\
                                  \e_2^{r_1} & \cdots & \e_2^{r_{n-1}} & 1 \\
                                  \vdots & \ddots & \vdots & 1 \\
                                  \e_n^{r_1} & \cdots & \e_n^{r_{n-1}} & 1 \\
                                \end{array}
                              \right).
$$This formula shows that the original landscape function on an $(N^2-1)$-dimensional group manifold $\G$ can be reduced to a function of $n-1$ independent variables $(\e_1,\cdots,\e_{n-1})$ on the $(n-1)$-tori $T^{n-1}$, thereby all $U\in\G$ corresponding to a critical point $(\e_1,\cdots,\e_{n-1})\in T^{n-1}$ form a subset of the critical submanifold with the same $J$ value.

As a direct application of the above formula, the case of $\SU(2)$ can be computed. Suppose that $\G$ carries a UIR $\D_j$ of $\SU(2)$, then the landscape function is
\begin{equation}\label{su2e1e2}
J_{\D_j}(\e_1,\e_2)=(2j+1)^{-1}\frac{\e_1^{2j+1}-\e_2^{2j+1}}{\e_1-\e_2}
\end{equation}according to (\ref{xi}). Let $\e_1=e^{i\beta}$ and $\e_2=e^{-i\beta}$, where $\beta\in[0,2\pi)$, then (\ref{su2e1e2}) can be reduced to a single variable function
\begin{equation}\label{J(beta)}
J_{\D_j}(\beta)=\Big|\frac{\sin\left[(2j+1)\beta\right]}{(2j+1)\sin\beta}\Big|,
\end{equation}whose critical points $\beta_1,\beta_2,\ldots$ characterize the critical topology.

Besides $\SU(2)$, the $\SU(3)$ case is relatively simple, but a universal solution is still not available for the critical topology of all its UIRs. However, as the corresponding character formula involves only two independent variables, it is possible to solve for the critical topology of low dimensional UIRs, and, numerically, it is easy to visualize the landscape in a 3D image for any individual UIR (labeled by two integers $r_1>r_2$). The dimension of the $(r_1,r_2)$ UIR is given by $N_{r_1,r_2}=\frac{r_1r_2(r_1-r_2)}{2}$. In the following, we will use $\SU(2)$ and $\SU(3)$ examples to study the impact of the symmetry group, the UIR dimensionality and the influence of realizability of the target gate upon the landscape topology. The case $\SU(n)$ for $n>3$ will not be discussed in this paper.

\subsection{The dimensionality of UIR}
Consider the $\SU(2)$ case with $W$ being realizable. Fig. 1 shows two examples with $j=3$ and $j=7/2$ for $\beta\in[0,\frac{\pi}{2}]$ (the remaining part $\beta\in[\frac{\pi}{2},2\pi]$ is repeating and thus omitted) according to the formula (\ref{J(beta)}), respectively,
corresponding to dimensions $N=7$ and $N=8$. Both landscapes
possess four local maxima, and their local minima
are all of value zero. No saddle points are found. There are three traps on the landscape, and the results obtained from the two examples can be generalized as below:
\begin{theorem}
Denote $\lfloor j\rfloor $ as the largest integer that is no greater than $j$. The
control landscape $J_{\D_j}$ has $\lfloor j \rfloor$ local
suboptima.
\end{theorem}

The systems that carry higher-dimensional UIRs (i.e., with a larger $j$ in $\mathcal{D}_j$) are less controllable because the number of degrees of freedom is always 3 (versus the total number $(2j+1)^2$ in $\U(2j+1)$). Thus, we can see from the Theorem that the search for an optimal control is more likely to be trapped on such landscapes because there tends to be a larger of number of traps. Moreover, the convergence region of the global optimum $U=\mathbb{I}$ can be identified as $\beta\in[0,\frac{\pi}{2j+1}]$ according to (\ref{J(beta)}), which also shrinks when $j$ increases, leaving the system more likely to be trapped by some of the $\lfloor j\rfloor-1$ local suboptima outside the region. Therefore, the loss of controllability leads to an increase of the ruggedness of the landscape.

\begin{figure}[h]
\centerline{
\includegraphics[width=3in,height=2in]{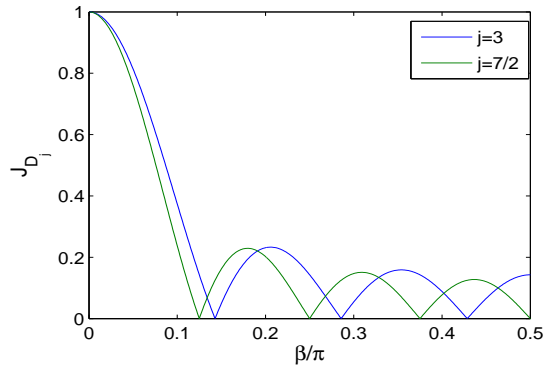}
} \caption{The character functions for $\D_3$ and $\D_{7/2}$
UIRs.}\label{D3D4}
\end{figure}

\subsection{The realizability of the target transformation}

\begin{figure}[h]
\centerline{
\includegraphics[width=5in,height=2.5in]{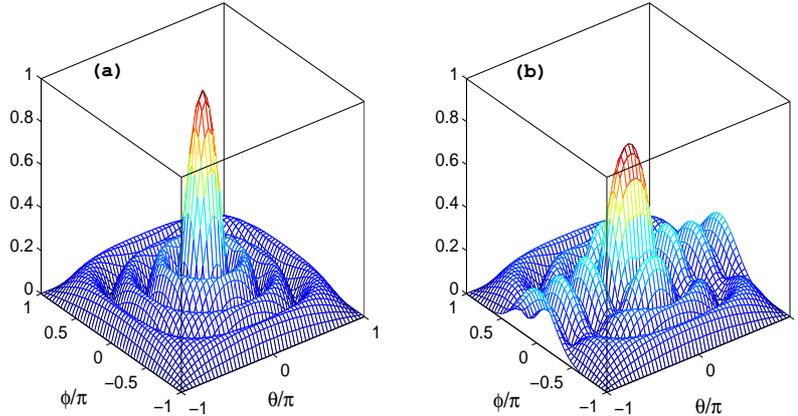}
} \caption{Two-dimensional illustration of the landscape topology of
$J(U)=N^{-1}|\tr(W^\dag U)|$ with (a) $W=\mathbb{I}_8\in \SU(2)$; (b)
$W=diag\{-1,1,\cdots,1\}\notin \SU(2)$.}\label{d4d5}
\end{figure}

Consider again the dynamic symmetry $\SU(2)$. If the target gate is not realizable, i.e., $W\notin\G$, then the dynamic symmetry group $\G$ is not invariant under the transformation $\tilde U=W^\dag U$, and hence the approach used above for realizable $W$ is no longer applicable. In this case, the landscape function cannot be reduced to having a single characterizing parameter, and we have to analyze it on the original three-dimensional group manifold. Here we adopt the well-know Euler decomposition
$$U(\psi_1,\phi,\psi_2)=e^{\psi_1 L_z}e^{\theta L_x}e^{\psi_2 L_z},$$
which parameterizes the landscape function by $(\theta,\psi_1,\psi_2)$. To facilitate visualization of the results, we consider the special case that the target gate $W$ commutates with $L_z$, i.e., $[W,L_z]=0$, for which the landscape function reduces to the following two-variable form:
\begin{eqnarray*}
J(U) &=& N^{-1}|\tr\{W^\dag e^{\psi_1 L_z}e^{\theta L_x}e^{\psi_2 L_z}\}| \\
 &=& N^{-1}|\tr\{W^\dag e^{(\psi_1+\psi_2) L_z}e^{\theta L_x}\}|\triangleq J(\theta,\phi),
\end{eqnarray*}where $\phi = \psi_1+\psi_2$. For example, suppose that $L_z$ is diagonal under some chosen UIR basis, then the landscape function can be reduced to a two-variable function if the target gate is a diagonal $N$-dimensional unitary matrix, whose diagonal elements affect the landscape topology. For such a class of unrealizable target gates, one can easily depict the two-dimensional landscape topology as exemplified in Fig.\ref{d4d5}. Compared to the case of a realizable $W$, the landscape hight (i.e., the maximal fidelity) decreases to $J_{\max}= 0.8$ due to the inability of perfectly realizing the gate $W$. In addition, the
landscape with realizable $W$ possesses three critical submanifolds (owing to the intrinsic dependence upon a single variable), while the landscape with an unreachable $W$ is much more rugged, implying
that having an unattainable target state not only degrades the maximal achievable fidelity but also increases the complexity of the landscape topology. Consequently, local search algorithms (e.g., gradient based methods) on such a control landscape likely will be less effective.

\subsection{Comparison between different dynamic symmetry groups}
\begin{figure}[h]
\centerline{
\includegraphics[width=5in,height=5in]{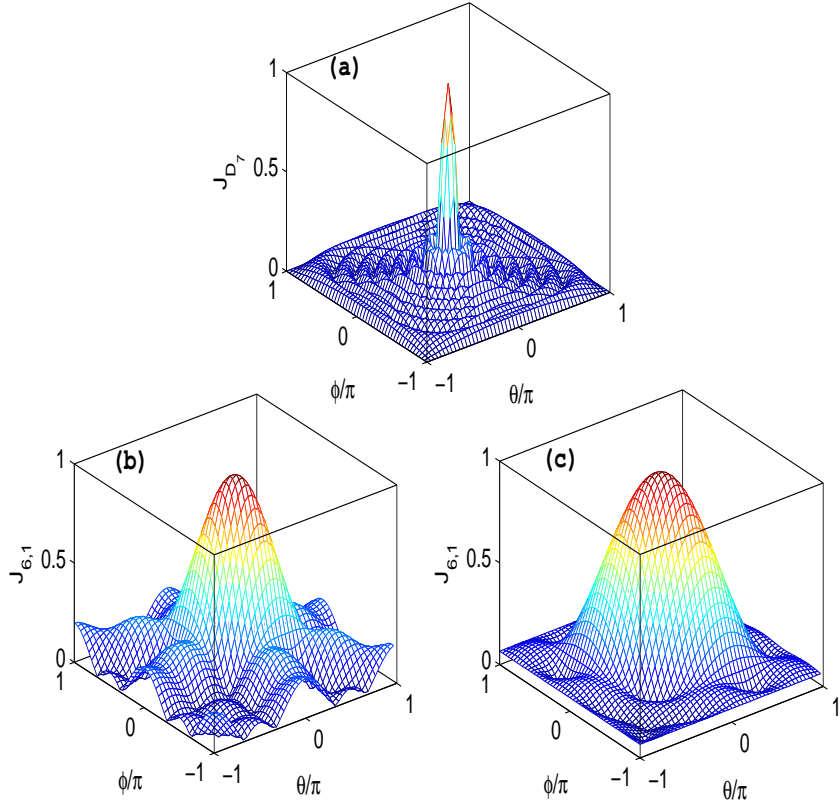}%{J42_3D}
} \caption{The gate fidelity landscape topology for $15$-dimensional systems under $\SU(2)$ and $\SU(3)$ dynamic symmetries: (a) the landscape with $\SU(2)$ symmetry with UIR $\mathcal{D}_7$; (b) the landscape with $\SU(3)$ symmetry with UIR labeled by $(6,1)$; (c) the landscape with $\SU(3)$ symmetry with UIR labeled by $(5,2)$. Their landscape ruggedness is in decreasing order.}\label{J42}
\end{figure}

To compare the influence of different dynamic symmetry groups, we choose the $(5,2)$ and $(6,1)$ UIRs of $\SU(3)$ and the UIR $\mathcal{D}_7$ of $\SU(2)$, whose dimensions are all fifteen (i.e., corresponding to different dynamic symmetries in the same physical system). We directly show the 3D landscape profiles in Fig. \ref{J42}. The landscape topology is less rugged for the $\SU(3)$ case, as all critical points are isolated. Moreover, the region of attraction for the global maximum is rather large, while that of the $\SU(2)$ symmetry is small, making its peak very sharp. This shows from another perspective that the enhanced controllability (corresponding to a larger dynamic symmetry group) helps to improve the search efficiency for global optimal controls by smoothing the landscape topology. It is also interesting that the landscapes of different UIRs of the $\SU(3)$ symmetry group can be distinct (i.e., the UIR $(5,2)$ landscape is less rugged) even when their dimensions are identical.

\section{Discussion}
The present study on the landscape of uncontrollable systems with $\SU(2)$ and $\SU(3)$ dynamic symmetries show that controllability plays an important role in maximizing unitary gate fidelity, whether or not the desired target is attainable. The loss
of controllability can lead to false traps over the
corresponding restricted kinematic control landscape on the dynamic symmetry group. The ruggedness generally increases when
controllability is weakened, i.e., either with a smaller dynamic symmetry group or an unattainable target unitary transformation, making the search for global optimal
controls much more difficult.

The details of the landscape topology depend on the particular structure of the dynamic symmetry group and the unitary representation it carries (see \cite{Polack2009} for a classification of such systems). Uncontrollable systems with dynamic symmetry generally possess rugged control
landscapes, but this may not always be the case. For example, the gate fidelity landscape with $W$ realizable and the following reachable sets in $\U(N)$ were found (or can be easily proved) to be devoid of local suboptima:
\begin{enumerate}
  \item Symmetry matrices in $\U(N)$ \cite{HsiehWu2010};
  \item Self-dual matrices  in $\U(N)$ \cite{HsiehWu2010};
  \item Symplectic matrices (symplectic group) \cite{WuRaj2010};
  \item Real orthogonal matrices (orthogonal group) in $\U(N)$.
\end{enumerate}
The landscape structure for the observable control problem \cite{Rabitz2005} also becomes rugged when the system is under dynamic symmetry. The corresponding the critical topology structure is dependent on the spectra of the system density matrix and the observable, and will be more complicated than the case of unitary fidelity landscape. This important problem will be explored in future studies.

\begin{acknowledgments}
The authors acknowledge support from the NSF. RBW also acknowledges support from NSFC (Grant No. 60904034).
\end{acknowledgments}

%\bibliographystyle{plain}
%\bibliography{TrapRef}

\end{document}